\documentclass[final,5p,times,twocolumn]{elsarticle}

\usepackage{lineno,hyperref}
\modulolinenumbers[5]

\usepackage{graphicx,url,twoopt,natbib}

\usepackage{siunitx}
\usepackage{color}
\usepackage{booktabs}

\usepackage{float}
\usepackage[labelformat=simple]{subfig}

\usepackage{multirow}

\makeatletter
\newcommand{\bibnote}[2]{\global\@namedef{#1note}{#2}}
\newcommand{\biblink}[2]{\global\@namedef{#1link}{#2}}
\makeatother

\begin{document}
\begin{frontmatter}

\title{A Study of background for IXPE}
\author[1]{F. Xie\corref{cor1}}
\ead{fei.xie@inaf.it}

\author[1]{R. Ferrazzoli}
\author[1]{ P. Soffitta}
\author[1]{ S. Fabiani}
\author[1]{ E. Costa}
\author[1]{ F. Muleri}
\author[1]{ A. Di Marco}

\address[1]{INAF-IAPS, via del Fosso del Cavaliere 100, I-00133 Roma, Italy}
\cortext[cor1]{Corresponding author}

\begin{abstract}
Focal plane X-ray polarimetry is intended for relatively bright sources with a negligible impact of background. However this might not be always possible for IXPE (Imaging X-ray Polarimetry Explorer) when observing faint extended sources like supernova remnants. We present for the first time the expected background of IXPE by Monte Carlo simulation and its impact on real observations of point and extended X-ray sources.
The simulation of background has been performed by Monte Carlo based on GEANT4 framework. The spacecraft and the detector units have been modeled, and the expected background components in IXPE orbital environment have been evaluated. We studied different background rejection techniques based on the analysis of the tracks collected by the Gas Pixel Detectors on board IXPE.
The estimated background is about 2.9 times larger than the requirement, yet it is still negligible when observing point like sources. Albeit small, the impact on supernova remnants indicates the need for a background subtraction for the observation of the extended sources.
\end{abstract}


\end{frontmatter}

%\linenumbers

%%%%%%%%%%%%%%%%%%%%%%%%%%%%%%%%%%%%%%%%%%%%%%%%%%%%%%%%%%%%%%%%%%%%%%%%%%%%
\section{Introduction}     \label{sec:intro}
%%%%%%%%%%%%%%%%%%%%%%%%%%%%%%%%%%%%%%%%%%%%%%%%%%%%%%%%%%%%%%%%%%%%%%%%%%%%

The Imaging X-ray Polarimetry Explorer (IXPE) is a NASA Astrophysics Small Explorer (SMEX) mission, selected in January 2017.
The launch is scheduled for 2021, by means of a Falcon 9 rocket that will deliver IXPE into an equatorial orbit with an altitude of 600 km.
IXPE is dedicated to X-ray polarimetry between 2 and 8 keV with photoelectric polarimeters based on the Gas Pixel Detector (GPD)  (\citep{2001Natur.411..662C}, \citep{2006NIMPA.566..552B}). The main scientific goals will be reached through the observations of known bright X-ray sources of different classes, including neutron stars, black holes, supernovae remnants (SNR), active galactic nuclei (AGN). 
X-ray polarimetry promises to evolutionize the knowledge of the X-ray sources.

Since X-rays from Scorpius X-1 have been detected in 1962 (\citep{1962PhRvL...9..439G}), many space missions have been approved for X-ray spectroscopy, imaging, and timing, unfortunately only very few measurements have been conducted on X-ray polarization.
In 1971, an Aerobee-350 sounding rocket was launched with two types of X-ray polarimeters onboard, a Thomson scattering polarimeter made of metallic lithium and a Bragg diffraction polarimeter made of graphite crystals. Both instruments used proportional counters as detectors. 
Limited by the observation time and unexpected high background level, only polarization of 15.4\% $\pm$ 5.2\% at a position angle of \ang{156} $\pm$ \ang{10} from the Crab nebula was measured with a statistical confidence level of 99.7\% (\citep{1972ApJ...174L...1N}). 
Later in 1975, the Orbiting Solar Observatory (OSO-8) satellite with two graphite polarimeters onboard was launched. A more precise polarization measurement of the X-ray flux from the Crab nebula has been reported in \citep{1976ApJ...208L.125W}, to be 16.1\% $\pm$ 1.4\% at \ang{160.2} $\pm$ \ang{2.6}, with a 10 standard deviation significance in six days observation.
Furthermore, \citep{1978ApJ...220L.117W} reported with the same polarimeter the polarization measurement of the Crab nebula without pulsar contamination, resulting on a polarization of 19.22\% $\pm$ 0.92\% at angle of \ang{155.79} $\pm$ \ang{1.37}, by combining results at 2.6~keV and 5.2~keV.
All these results confirmed the synchrotron radiation origin of the X-ray emission from the Crab nebula.
Since then, no polarimeter ever has been flown in soft X-ray range, till 29th October 2018, when a CubeSat mission named Polarimeter Light (PolarLight), based on the GPD technique but with a collimator, was launched into a Sun-synchronous orbit~(\citep{2019ExA....47..225F}). 
After more than one year operating in space, PolarLight has proven the operation of GPD. Due to the design limitations typical of cubeSat (small area, no optics), PolarLight has scientific capability only on very bright sources such as the Crab~(\citep{2020NatAs...4..511F}).

IXPE is foreseen to improve the sensitivity over OSO-8 by two orders of magnitude.
It is equipped with three identical mirror modules and three identical detector units (DUs) located at their focus. 
The heart of the telescope, the polarization-sensitive detector GPD, has been described in detail in~\citep{2001Natur.411..662C}, \citep{2013NIMPA.720..173B}, \citep{2016SPIE.9905E..17W}, \citep{2017SPIE10397E..0IS}, and will only be reviewed here briefly.
GPD exploits the photoelectric effect in the gas. The emission directions of the photoelectrons are related to the polarization of the incident X-ray photons. 
The GPD gas cell is filled with Dimethyl Ether (DME) gas at 0.8~atm, which is closed on the top by the Titanium frame and the Beryllium window. An electric field parallel to the optical axis drifts the primary electrons produced by the photoelectron in the gas to the Gas Electron Multiplier (GEM), which is also the bottom of the active volume. Secondary electrons, generated by the GEM, are collected by a pixellated plane at the top of an Application Specific Integrated Circuit (ASIC). The side of the gas cell are Macor spacers. 
The gas cell has dimension of $60\times60\times10$~mm$^3$, but the active volume, defined as the region where charge generated by an energy deposit can be read out by the ASIC pixels, is only the central $15\times15\times10$~mm$^3$ volume. The ASIC has $352\times300$ pixels arranged in a hexagonal pattern with 50 $\mu$m pitch, and it provides the fine spatial resolution required to resolve the photoelectron track.

The standard parameter to express the sensitivity to polarization is the Minimum Detectable Polarization (MDP) at a confidence level of 99\%. It is defined as (\citep{2010SPIE.7732E..0EW}):
\begin{equation}
MDP_{99} = \frac{4.29}{\mu S} \sqrt{\frac{S+B}{T}}
\label{equ:mdp}
\end{equation}
Where S and B are the count rates of the source and background, T is the total exposure time and $\mu$ is the modulation factor representing the amplitude of the response to 100\% polarized beam.
Reducing the background is vital for achieving a high sensitivity.
For bright point sources ($S \gg B$), background is negligible in practical cases as we will demonstrate below.
But for extended sources, even the brightest ones, like pulsar wind nebulae, SNRs, and AGN large scale jets, for which IXPE will perform X-ray polarimetric imaging for the first time, the background modeling must be treated with care.

In this paper, we evaluate the background based on Monte Carlo simulation, to estimate the in-flight background level in a detailed way. By understanding the characteristics of the background and their impact on the statistics of the measurement, methods for background rejection are proposed. 
This paper is organized as it follows: Section~\ref{sec:bkgsim} presents the details of the simulations, and Section~\ref{sec:results} presents results from the simulation as well as the developing of the rejection methods. Section~\ref{sec:discussion} discusses the impact of background for point and extended sources, and Section~\ref{sec:conclusion} presents the conclusion of the work.

%%%%%%%%%%%%%%%%%%%%%%%%%%%%%%%%%%%%%%%%%%%%%%%%%%%%%%%%%%%%%%%%%%%%%%%%%%%%
\section{Background simulation}
\label{sec:bkgsim}
Background simulation are developed based on the Geant4 framework (\citep{2003NIMPA.506..250A}), a toolkit for Monte Carlo simulations.
By tracking the particles interacting with the detector, the origin and the phenomenology of the background are understood.
To perform the simulation, the geometric model of the spacecraft, the input background spectra of the orbit environment and the physics model of the interaction are needed. 
Then, mimicking the data processing and analysis of real data, selections are applied in order to determine the background rates.
Geant4 \textcolor{black}{version 10.03.p01\footnote{\url{http://geant4.web.cern.ch/support/download_archive?page=1}}} has been used in this work.

\subsection{Geometric model}
The geometric model is an important ingredient for Monte Carlo simulation. In IXPE simulator, the construction of the geometric model is based on the latest design (\citep{2016SPIE.9905E..17W}) and implemented in two steps. The first step is the construction of the core payload GPD, which is implemented in the \textrm{ixpesim} software developed by the IXPE Italian Team. From the top to the bottom there are 50~$\mu$m Beryllium window with 53~nm Aluminum coating layer, 10~mm DME gas and 50~$\mu$m GEM with 5~$\mu$m Copper coating on both sides. 
Gas is enclosed at their sides by the MACOR (a glass ceramic mainly composed of silica SiO$_2$ and various oxides, including MgO, Al$_2$O$_3$ K$_2$O and B$_2$O$_3$ etc.) spacer. This is the basic setup in the laboratory. 
Both simulations and measurements show that when a photon is absorbed near the surface separating the DME gas cell and the passive materials that seal it, in many cases the charges produced by ionization processes are partially generated in the gas, where they are detected, and partially in the Be window or in the GEM where they are lost. Further events could be generated in the gas though the initial photons are already lost in the passive materials. 
These kind of events usually result in a left tail in addition to the Gaussian profile for a monochromatic beam and have relatively smaller track size. But this also requires that the materials contiguous to the gas cell are described very carefully in the mass model.
\begin{figure}[hbtp]
\centering
\includegraphics[width=0.65\columnwidth]{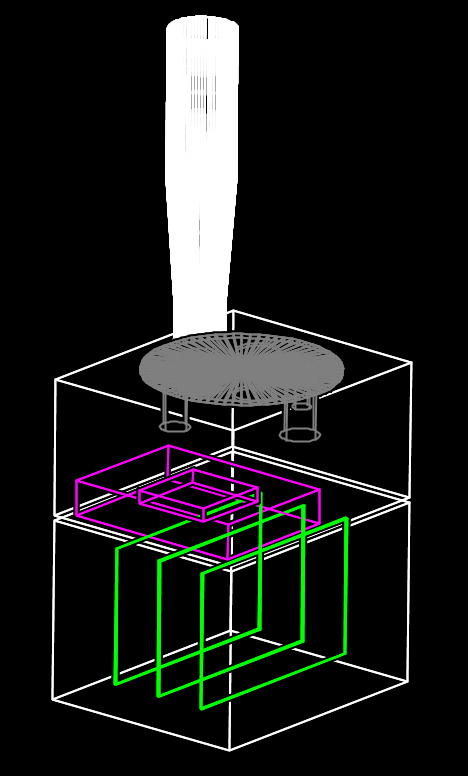}
\caption{A sketch of the Geant4 mass model of one detector unit, collimator (white), calibration wheel and calibration sources (grey), GPD (magenta) and PCBs (green) are presented. Structural details are hidden for a clear illustration.}
\label{fig:g4du}
\end{figure}

To evaluate the background in a realistic configuration, we built on top of the first step a model of the whole satellite.
Materials around the detector are essential for background simulation. Background particles mainly come from the region outside the field of view (FoV). 
They come across the material around the detector producing secondary particles. Some interact with the detector leaving a stream of charge in the gas which are eventually detected by the GPD.
For each DU, the calibration wheels, the collimator, the detector shielding box and the back-end electronic box are implemented, as shown in Fig.~\ref{fig:g4du}. Three identical DUs are located on the top deck of the platform with $120^\circ$ clocking of one DU with respect to the others.
The mirror modules, the deployable boom and the spacecraft are constructed as well, though with less details.
In order to speed up the simulation, the mirrors and boom are eventually removed from the mass model, therefore an efficient spatial sampling of the primary source particles has been applied. The test sample shows that the lack of mirrors and boom have no influence on the background rate as they are far away from the sensitive detector, except for the cosmic X-ray background (CXB). The materials around the detectors stop the low energy CXB photons well. 
Without the mirrors, the background rate induced by the CXB increases significantly because the active volume of the detector is now directly exposed to the sky through the aperture of the stray-light collimator. A cut on the incidence angle has been applied on the CXB background data, by assuming that mirrors stop the CXB background totally. The CXB reflected by the mirrors is not part of the present work, as it gives a minor contribution to the background in the IXPE energy band.

%%%%%%%%%%%%%%%%%%%%%%%%%%%%%%%%%%%%%%%%%%
\subsection{Space radiation environment}
\label{subsec:spaceEnv}
\begin{figure}
\centering
\includegraphics[width=\columnwidth]{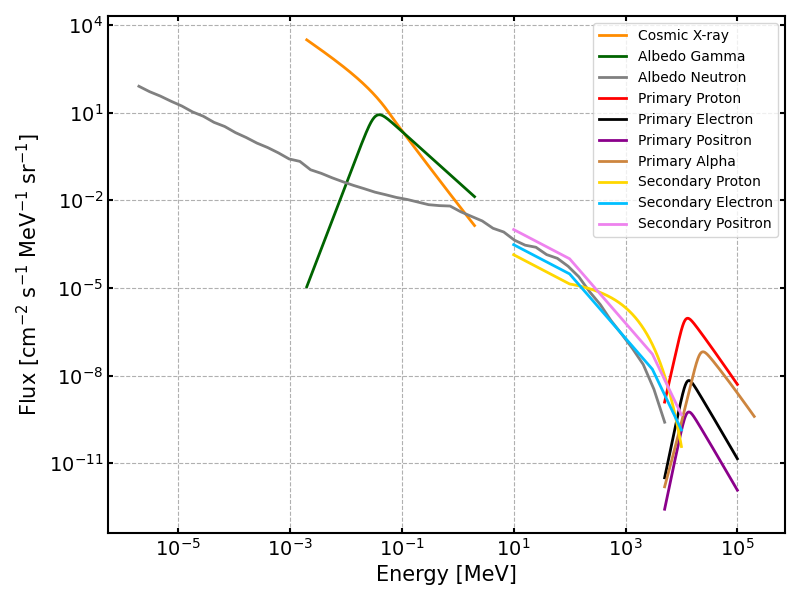}
\caption{The spectra of the background components expected in the IXPE orbital space environment.}
\label{fig:bkgspec}
\end{figure}

IXPE will be launched into an equatorial orbit with an altitude of 600~km, which minimizes the detector background and optimizes the observing efficiency by minimizing the passage in the South Atlantic Anomaly (SAA), achieving a minimum duty cycle of 60\% from the 94.6-minute orbital period.
In the radiation environment outside the SAA, the background sources which need to be considered are primary (protons, electrons, positrons and alpha particles) and secondary (protons, electrons, positrons) cosmic rays. Primary cosmic rays are accelerated by celestial sources and travel through the galaxy before reaching the Earth. The dominant component ($\sim$90\%) are protons. 
When primary particles impinge on the top of the atmosphere and interact with residual gas molecules, showers of secondary cosmic rays are produced and some of these eventually go upward and escape the atmosphere (\citep{2004ApJ...614.1113M}). 
We also study other albedo components generated in the atmosphere, such as gamma-rays, neutrinos and the extragalactic CXB. The latter is usually the dominant background component for large FoV X-ray telescopes, but this is not the case for IXPE.
The input spectra are originally derived from simulations for the LOFT mission proposal (\citep{2013ExA....36..451C}) which assumes a low Earth orbit with an altitude of 600 km and an inclination of $5^\circ$, then adapted for the XIPE (X-ray Imaging Polarimetry Explorer) mission phase A study with $0^\circ$ inclination baseline design at similar altitude, and eventually used for IXPE in this work.
The input spectra of these sources are detailed in Fig.~\ref{fig:bkgspec}.

\subsection{Physics model}
In order to fully describe the interactions in the space environment, electromagnetic, hadronic, and decay processes are included. For the electromagnetic interaction, a reference physics list provided by Geant4, ``G4EmLivermorePolarizedPhysics'' \footnote{\url{https://geant4.web.cern.ch/node/1619}}, has been chosen as the starting point. It is a physics list recommended for low energy electromagnetic processes, describing the interactions of electrons and photons with matter down to about 250~eV, with the polarized gamma models included. After the cross check with laboratory data, physics processes and parameters are fine tuned to fit the specific case of IXPE. 
For example, when considering the photoelectric effect, the build-in ``G4LivermorePolarizedPhotoElectricModel" has been replaced by ``G4LivermorePolarizedPhotoElectricGDModel". Because this model is optimized for measuring linearly polarized X-rays in the energy range of few keV, and it properly takes the directions of the photoelectrons into account (\citep{2006NIMPA.566..590D}).
The other physics processes are covered by Shielding physics list, which is another reference list for space missions. Different production thresholds for secondary are applied for different regions. A default value of 0.7~mm has been kept for the world, while 0.05~mm is applied for gas cell. This means the secondary particles are only generated when the kinetic energy is large enough for them to travel 0.05~mm in the gas.

\subsection{Simulation logic}
\label{subsec:logic}
When a particle arrives at the telescope, all the interactions are tracked, including the secondaries generated by this particle that we name primary. 
For clarity, the terms primary and secondary used here are not the same as the definitions in~\ref{subsec:spaceEnv}. 
Referred to the technique of simulation, each individual incident particle is a primary, it could be any kind of particle we are interested in. One primary particle, especially an energetic one, may generate a bunch of secondary particles through interactions with the mass model built inside the simulator. These secondaries have possibilities to be detected and become background events.
Considering the storage space, only the energy deposited inside the gas is recorded. 
If the particle is a photon, it either passes through the gas without any interaction, or scatters on the electrons, or excites the inner-shell electrons to free with kinetic energy of the difference between the photon energy and the binding energy.
If it is a charged particle, it ionizes the gas along the path or scatters with the nuclei, untill running out of the kinetic energy or exiting the sensitive area. 
Along the path inside the gas cell, starting from the point of entry up to either the zero-kinetic-energy or the point of exit, hundreds of electrons are generated and their positions are recorded. 
Electrons are drifted to the corresponding holes of the GEM, meanwhile the transverse diffusion, the Fano factor and the absorption attachment (\citep{2017SPIE10397E..0HF}) of the gas are considered. An analytic multiplication is applied on the number of electrons for each hole of the GEM, to represent the avalanche multiplication of charges.
By taking into account the front-end gain, the full-scale voltage range and the resolution of the analog to digital converter (ADC), the Monte Carlo digitize the charge (number of electrons) into the ADC value and projects them to the corresponding hexagonally patterned ASIC pixels. The projected charge distribution on the pixelated ASIC plane is defined as the track.

As for the real detector, an event is read out only when it fulfills two conditions: 
(1) At least one mini-cluster (2$\times$2 pixels) has the energy deposit above the threshold; 
(2) The number of pixels for the region of interest (ROI) -- window size, is in between of 30 to 5000. 
The ROI for each event is defined as a rectangular area containing all the triggered mini-clusters plus a margin of 10 pixels. Both the trigger threshold and range of the window size are adjustable. 

After readout, the analysis pipeline applies the reconstruction algorithm on the track to get the ejection direction of the photoelectron, which carries the memory of the source polarization. Inside the algorithm, zero suppression of the noise,  pixels clustering, first and second moment analysis are implemented, more details are discussed in \citep{2003SPIE.4843..383B}, \citep{NiccoloPhdThesis}. 
In addition to photoelectron direction, reconstruction algorithm derives a number of relevant track properties, which are important for background rejection methods development,  and they will be explained in Section~\ref{subsec:bkgreject}.

%%%%%%%%%%%%%%%%%%%%%%%%%%%%%%%%%%%%%%%%%%%%%%%%%%%%%%%%%%%%%%%%%%%%%%%%%%%

%%%%%%%%%%%%%%%%%%%%%%%%%%%%%%%%%%%%%%%%%%
\begin{table*}
\caption{Count rates for all the background components.} 
\centering
\begin{tabular}{cccc}
\toprule
\textbf{Component}	& \textbf{Rate in total [s$^{-1}$]}& \textbf{Rate in 2--8 keV [s$^{-1}$]}& \textbf{Reject efficiency [\%]}\\
\midrule
Cosmic X-ray		& 	3.19E-03	&	1.73E-03	&	45.76	\\
Albedo Gamma		& 	3.39E-03	&	1.24E-03	&	63.48	\\
Albedo Neutron		& 	1.14E-03	&	2.97E-04	&	74.01	\\
Primary Proton		& 	9.77E-02	&	3.16E-02	&	67.65	\\ 
Primary Electron	& 	8.67E-04	&	2.39E-04	&	72.43	\\
Primary Positron	& 	7.45E-05	&	1.91E-05	&	74.34	\\
Primary Alpha		& 	3.03E-02	&	1.09E-02	&	63.95	\\
Secondary Proton	& 	4.50E-02	&	1.41E-02	&	68.63	\\
Secondary Electron	& 	4.16E-02	&	1.11E-02	&	73.35	\\
Secondary Positron	& 	1.32E-01	&	3.36E-02	&	74.61	\\
\midrule
Total				&	3.56E-01	&	1.05E-01	&	70.51	\\
\bottomrule
\end{tabular}
\label{table:bkgrate}
\end{table*}

\section{Results}  \label{sec:results}
\subsection{Overview of the background level}
We simulated CXB, albedo gamma, albedo neutron, primary and secondary cosmic rays as mentioned above, with exposure long enough to achieve a sufficient statistic. The background rates of all the components are listed in Table~\ref{table:bkgrate}. 
The column `Rate in total' are the read out events fulfilled conditions introduced in Section~\ref{subsec:logic}, showing that the recorded events are mainly from the primary proton, primary alpha and the secondary cosmic rays. Photon-origin background including CXB and albedo gamma are not significant, while contribution from albedo neutron, primary electron and positron are negligible. 

An energy selection for the range between 2 and 8~keV cuts the total background rate down to $1.05\times10^{-1}$~counts\,s$^{-1}$, with a rejection efficiency of 70.5\%, as shown in Table~\ref{table:bkgrate}.
Photon-origin background have the lowest rejection efficiencies with this method. 

The scientific requirement of the background level is $4\times10^{-3}$~counts\,s$^{-1}$\,cm$^{-2}$ for one DU in 2 to 8~keV, which is estimated considering the most extended and faintest sources in the IXPE target list (X-rays reflected from the Sgr~B2 molecular clouds in the vicinity of the Galactic Center). Considering the geometric area of one DU of 2.25~cm$^2$, a background level of ($1.05\times10^{-1}$) / (2.25) = $4.66\times10^{-2}$~counts\,s$^{-1}$\,cm$^{-2}$ is the result with only energy selection. This is about 12 times higher than the requirement. An efficient background rejection method is needed.

%%%%%%%%%%%%%%%%%%%%%%%%%%%%%%%%%%%%%%%%%%
\subsection{Background rejection}
\label{subsec:bkgreject}

\begin{figure}[h]
\centering
\subfloat[]{
\includegraphics[width=0.95\columnwidth]{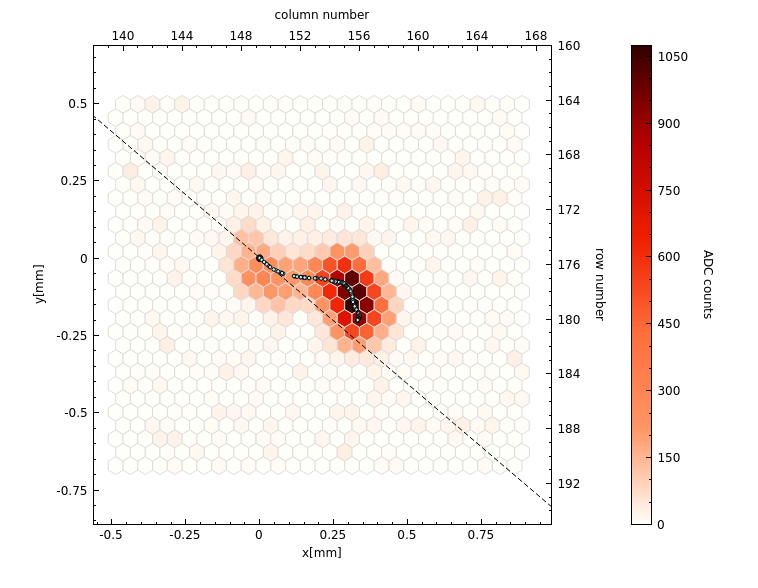}}
\hfill
\subfloat[]{
\includegraphics[width=0.95\columnwidth]{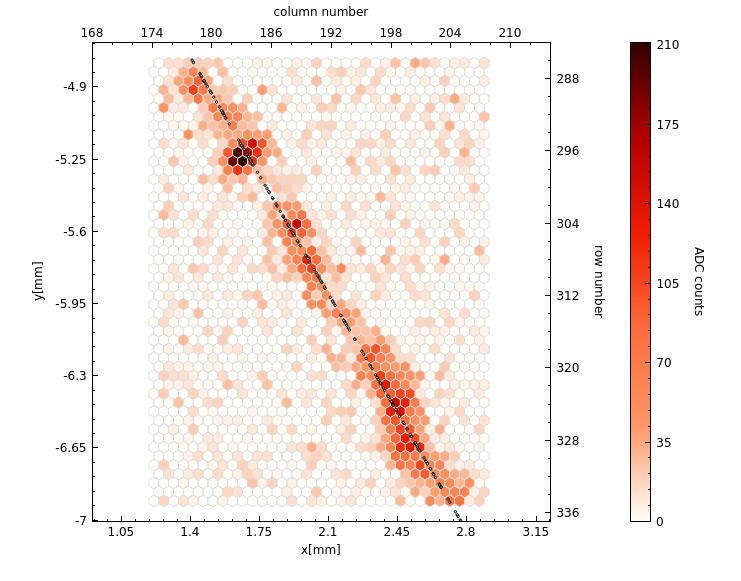}}
\captionsetup{width=1\linewidth}
\caption{Imaging of tracks with random electronic noises from the simulation. The dots represent the energy deposit 2D positions inside the gas recorded by Geant4.
	{\em (a)\/}: a track from a Fe$^{55}$ decay X-ray photon and dashed line is the photoelectron eject direction.
	{\em (b)\/}: a track from an energetic cosmic ray proton.} 
\label{fig:trackcmp}
\end{figure}

Background rejection methods are based on the fundamental differences between photoelectron and background tracks. A straightforward comparison is shown in Fig.~\ref{fig:trackcmp}, which shows two tracks with similar energy deposit in the detector but different origin. 
The case (a) is a classical photoelectron track from a 5.9~keV X-ray photon decayed from Fe$^{55}$. From the Bethe formula, the energy loss increases as the particle velocity is decreased. Therefore, photoelectron tracks shows the maximum charge density at its end, which is named Bragg peak. Dashed line presents the photoelectron eject direction. 
The case (b) shows, instead, a track generated by an energetic proton (tens of GeV). The charge particle exits the gas leaving a long, discontinuous string of electrons behind through ionizing.
Background tracks do not always look like Fig.~\ref{fig:trackcmp} (b), they may also be similar to photoelectron tracks, depending on the particle type, the kinetic energy, the incident direction and the interaction physics process. In the meanwhile, some photoelectron tracks also deviate from the ideal cases. The principle for background rejection is to remove background events as much as possible while keeping source photons. This is done by parametrizing the properties efficient in recognizing the background tracks. 

\subsubsection{Relevant properties for background rejection}
\label{subsubsec:propetries}
\begin{figure}
\centering
\includegraphics[width=\columnwidth]{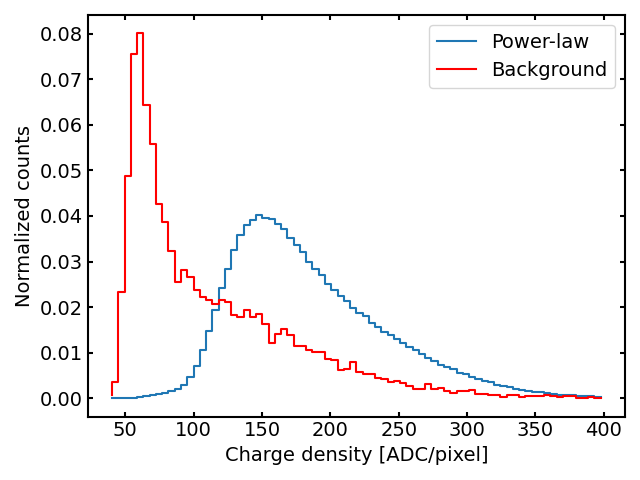}
\caption{Comparison of the charge density distributions expected from a power-law spectrum with the photon index of 2 (in blue) and the background including all the simulated components (in red).}
\label{fig:chargeDensity}
\end{figure}

We have studied most of the properties derived from the track analysis, for both genuine photoelectron tracks and background events. In this section, we introduce the beneficial parameters to be used for background rejection, with definitions listed as below:
\begin{enumerate}
\item \textbf{Pulse Invariant (PI)}\\
PI is the sum of the charge of the track, which is proportional to energy deposit. It is conventionally expressed in ADC channels and it is intended to be correct for the possible non-uniformity in the detector gain.
When calculating the background rate, only tracks with energy depositing in 2 to 8 keV, which is the IXPE energy range, are counted.
\item \textbf{Track size} \\
Track size is the number of pixels above the threshold in the main cluster, that is, in the largest group  of contiguous pixels of the event.
With the same energy deposit, background events usually leave larger track size than photoelectrons.
\item \textbf{Skewness} \\
Skewness, the third standardized moment, refers to the asymmetry of the energy distribution in the track along the major axis. 
The mean energy loss of a charged particle varies inversely with its energy. For a photoelectron of a few keV, at the very end of its path, the energy loss is progressively increasing toward the end point, forming a skewed track. On the contrary, for a background charged particle of the order of MeV or GeV,  the energy loss and therefore the ionization density is constant and the track has a low skewness.
For example, in Fig.~\ref{fig:trackcmp}, photoelectron track (a) is more skewed (asymmetric) than background track (b), and this is the usual case for them.
\item \textbf{Elongation} \\
Elongation is defined as $\sqrt{M2L/M2T}$, where M2L and M2T are the longitudinal and transverse second moments of the track, and their ratio refers to the eccentricity of the charge distribution. 
\item \textbf{Charge density} \\
Charge density is defined as the energy (PI) divided by track size, which is expected to be lower for background than photoelectron. For example, the relativistic background, known as minimum-ionizing particle (MIP), has energy losses about 2~MeV per\, g\,cm$^{-2}$ in light material (\citep{2000rdm..book.....K}), while the photoelectron energy density is about 10 times larger than that of a charge particle (\citep{2012SPIE.8443E..1FS}). 
A comparison is shown in Fig.~\ref{fig:chargeDensity}. The position of the peak from the background is smaller than that of a source simulated as a power-law spectrum with a photon index of 2. The significant deviation benefits background identification.
\item \textbf{Cluster number} \\
Cluster number is the number of clusters after applying clustering algorithm in the ROI.
For the same reason above, a long path length with low energy density from background particle is more likely discontinuous, therefore more than one cluster may be grouped by the clustering algorithm. On the contrary photoelectrons are mainly grouped as single cluster only.
\item \textbf{Border pixels}\\
Border pixels is the number of pixels in the track which are at the edge of the ASIC. Background entering in the gas from the side of the wall have a larger probability of leaving a track with pixels on the border.
\end{enumerate}

%%%%%%%%%%%%%%%%%%%%%%%%%%%%%%%%%%%%%%%%%%
\subsubsection{Development of background rejection}
Before applying background rejection we need to quantify the range of the parameters useful for this aim.
For each readout event, all of these parameters are derived from the track analysis. For millions of events, there will be the probability distribution of each parameter. We firstly determine the range of these parameters from the source simulation,  then apply the cuts to the background events, and finally reject the background events if the parameters are not in the accepted range. The parameters of tracks from the photoelectrons vary significantly with their kinetic energies. 
For example the track size of a 2~keV photon (see Fig.~\ref{fig:tracksize}) has a most probable track size of 45 pixels, while an 8~keV photon has a most probable track size of 126 pixels.
Separate such events into different energy bins helps to quantify the relevant parameters with a narrower range, therefore it is more efficient in recognizing the background tracks.
The following steps allow for developing the rejection methods:

\begin{figure}
\centering
\includegraphics[width=\columnwidth]{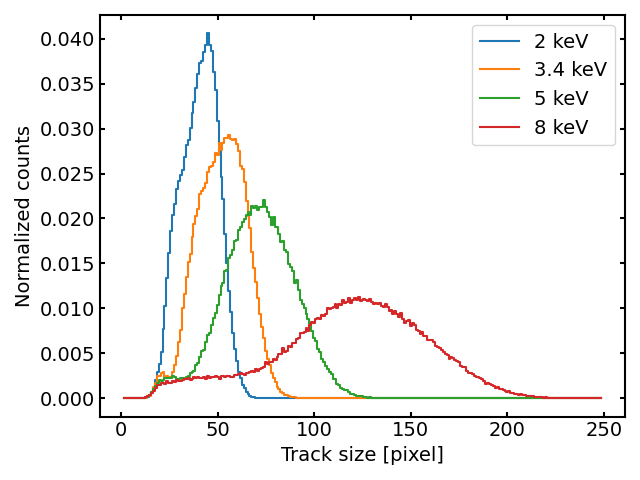}
\caption{Comparison of the track size distribution. Results from the monochromatic photons in four different energies are presented in different colors.}
\label{fig:tracksize}
\end{figure}

1) In consideration of the energy resolution of the GPD, we firstly divide the full energy range 2--8~keV into 3 energy bins, 2--3.4~keV, 3.4--5~keV and 5--8~keV (hereafter called Bin 1, Bin 2 and Bin 3). 
From the calibration, the energy resolution is about 18\% at 5.9~keV and at the other energies resolutions are described approximately as a function that scales as E$^{-1/2}$, therefore three energy bins are a reasonable assumption. 
When quantifying the parameters, an independent study will be conducted in each of these energy bins.

\begin{figure}
\centering
\includegraphics[width=\columnwidth]{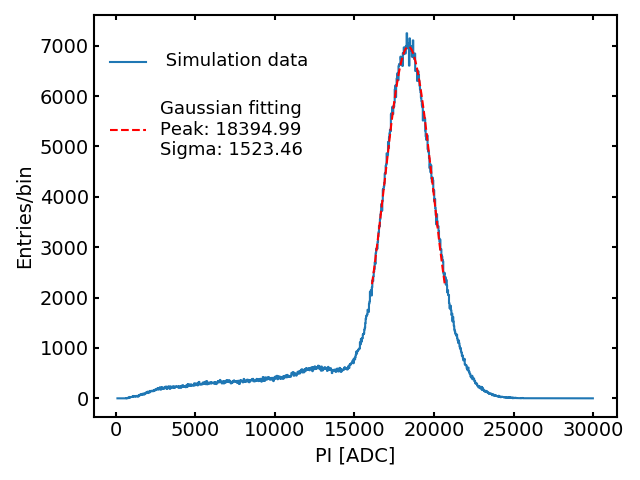}
\caption{PI distribution of 5.9~keV monochromatic photons, the red dashed line is the Gaussian fitting curve.}
\label{fig:fe55}
\end{figure}

2) The second step is to convert the boundaries of the chosen energy bins, i.e. 2~keV, 3.4~keV, 5~keV and  8~keV, to the corresponding ADC counts, which is the unit used in the simulations for the energy deposit and which is then used to apply events selection.
In our simulator, the energies of the readout events are given in ADC counts, for the purpose of treating the energy resolution and noise suppression properly, which is important in the step of reconstruction (track analysis). 
The correlations between the deposited energy and ADC counts are well calibrated to be in excellent agreement with the laboratory measurement.
Fig.~\ref{fig:fe55} shows the energy distribution in ADC counts from the simulation of 5.9~keV decay photons from the radioactive isotope Fe$^{55}$. The best Gaussian fitted peak value of 18395 ADC counts equal to 5.9~keV,  the resolution is defined as the ratio of the full width at half maximum (FWHM) to the peak value.
The connections between the bin-edge energies and the ADC counts are listed in Table~\ref{table:bkgadc}.
Once the gain has been set, the same correlation applies for both the simulation of source and background. 

\begin{table}[hbtp]
\caption{Connections between the bin-edge energies and the ADC counts, the peak and sigma are the values from the Gaussian fitting, resolution is the ratio of FWHM to the peak value.}
\centering
\begin{tabular}{cccc}
\toprule
\textbf{Energy [keV]}	& \textbf{Peak}& \textbf{Sigma}& \textbf{Resolution [\%]}\\
\midrule
2 		& 	6092		&	820		&	31.7	\\
3.4 		& 	10485	&	1058		&	23.7	\\
5 		& 	15532	&	1306		&	19.8	\\
8		& 	 24950	&	1696		&	16.0	\\
\bottomrule
\end{tabular}
\label{table:bkgadc}
\end{table}

\begin{figure}
\centering
\includegraphics[width=\columnwidth]{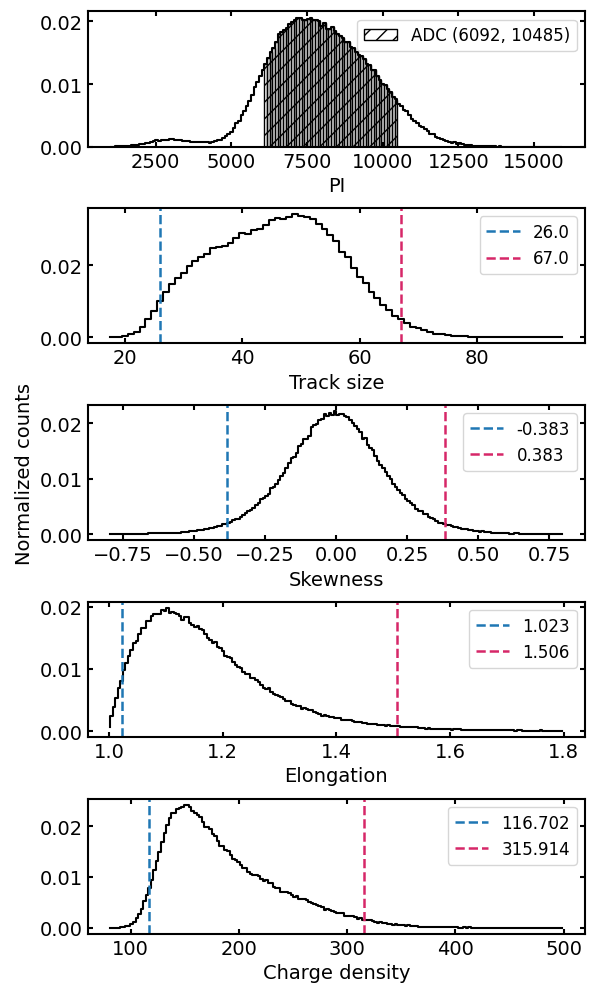}
\caption{Parameters distribution for flat-distributed photons in Bin 1. The hatched area in the top panel are the energy selection events. The head 2\% and tail 2\% are illustrated in blue and red lines.} 
\label{fig:getRejectPara}
\end{figure}

\begin{table*}
\caption{Parameters and the corresponding ranges applied for background rejection.}
\centering
\begin{tabular}{cccccc}
\toprule
\textbf{Energy bin}	& \textbf{PI}& \textbf{Track size}& \textbf{Skewness}& \textbf{Elongation} & \textbf{Charge density}\\
\midrule
Bin 1		& 	(6092, 10485)	&	(26, 67)	&	(-0.383, 0.383)	&	(1.023, 1.506)	&	(116.702, 315.914)\\
Bin 2		& 	(10485, 15532)	&	(34, 92)	&	(-0.620, 0.620)	&	(1.044, 2.414)	&	(140.186, 315.930)\\
Bin 3		& 	(15532, 24950)	&	(48, 156)	&	(-1.013, 1.010)	&	(1.112, 4.386)	&	(134.936, 386.334)\\
\bottomrule
\end{tabular}
\label{table:bkgpara}
\end{table*}

3) The next step is to simulate genuine photons with a flat distribution in each energy bin, and select the events in each bin according to the PI of the event.
We take the simulation for Bin 1 as an example illustrated here, the procedure for the other bins are the same.
A uniformly distributed photons in 2--3.4~keV are generated impinging on detector along the telescope optical axis. After applying the reconstruction software on the readout data, a series of parameters is determined. 
Firstly we apply the energy cuts on PI to select the good events. 
Even though all the readout events are originally from the photons in the energy range of 2--3.4~keV, in reality nothing about the original energy of the particles is known, but the deposited energy PI in ADC counts. From the previous step and correlations in Table~\ref{table:bkgadc}, only events with PI in between 6092 and 10485 ADC counts are taken into account in the case of Bin 1. As shown in the top panel of Fig.~\ref{fig:getRejectPara}, the black line presents all the read out events, while only the hatched area are the selected events, which will subsequently be used for the data analysis. 

4) The last step is to identify the parameters for an efficient background rejection and quantify their range.
Parameters have been studied by comparing the distribution difference between photoelectron tracks and background tracks. From all the possible parameters, we picked out the most efficient ones: track size,  skewness, elongation, charge density, cluster number, and border pixels (see their definitions in Section~\ref{subsubsec:propetries}).
The wider the range of the selected parameter, the less efficient the background rejection.
We fixed the accepted range of each parameter by removing 4\% of the events (2\% at the head and 2\% at the tail) in the distribution of that parameter for genuine event from photons in the chosen energy bin. This is shown in four bottom panels of Fig.~\ref{fig:getRejectPara}.
The black lines present the events after energy selection (hatched area from the top panel), blue and red lines define the two 2\% boundary from the distribution of the relevant parameters.
The quantified edges for all the three bins are listed in Table~\ref{table:bkgpara}.
We also request that the number of clusters is one and border pixels are zero for event acceptance.

\begin{figure}
\centering
\includegraphics[width=\columnwidth]{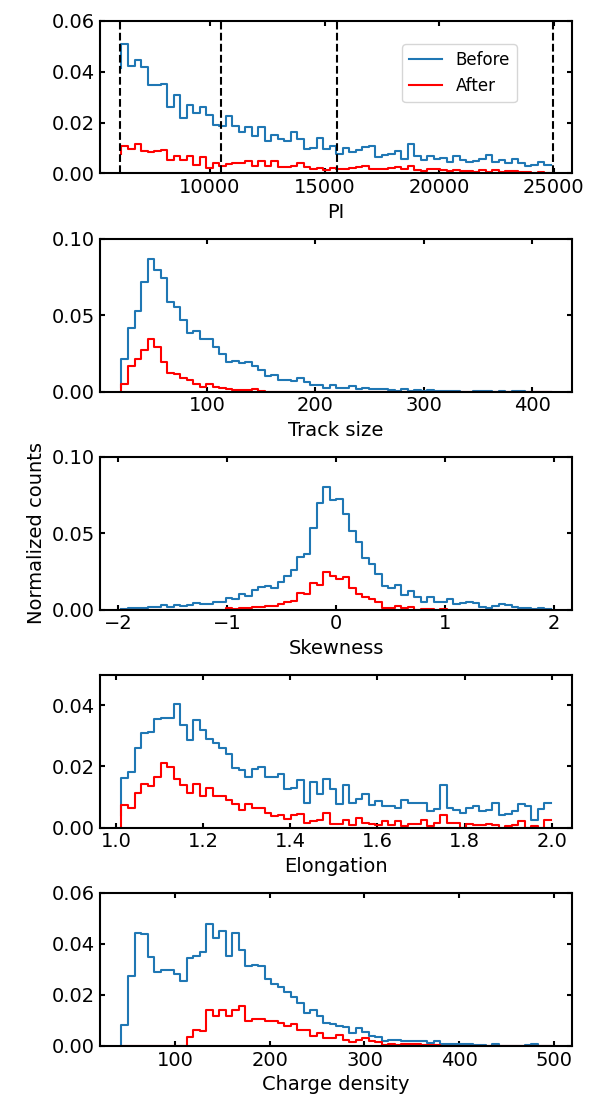}
\caption{Parameters distribution for background before (in blue) and after (in red) applying the rejection methods. Dashed lines in the top panel mark the edges of three energy bins.}
\label{fig:bkgspecreject}
\end{figure}

\subsubsection{Techniques of background rejection and impact on the signal}
The results of this study may be used to optimize the signal to noise ratio, namely the sensitivity, of IXPE. This is based on the rejection of some events  with criteria (often referred to as cuts) that are not foreseen during on ground activities, when the nature and amount of background are totally different. Beside studying these cuts following as guideline on the sensitivity, we must verify whether any systematics is introduced. For this activity simulations are needed. The eventual validation of these methods will be done with flight data but always checked for systematics by using the on-ground and in-flight calibrations.

In this section, we evaluate the residual background after applying the rejection methods introduced above.
We aim at a high rejection efficiency on background and at a low removal of true X-ray events from target source.
Furthermore, this method should not induce a bias on the polarization detection.
The following steps allow us to perform the verification.

1) We firstly apply the cuts to the background simulation data. 
The total background spectra before and after rejection are shown in Fig.~\ref{fig:bkgspecreject}, as well as the relevant parameters. More events from Bin 1 have been rejected, but still background is dominated by low energy events.
The residual background levels for all the components are listed in Table~\ref{table:bkgreject}. 
The total background reduces to $2.62\times10^{-2}$~counts\,s$^{-1}$, i.e. $1.16\times10^{-2}$~counts\,s$^{-1}$\,cm$^{-2}$ in 2--8 keV, with an extra rejection efficiency of 75.0\%, though it is still larger than the requirement by a factor of 2.9.
The main components of background events after applying the rejection methods are still the same. 
Compared to the rejection efficiency for the other components (except for the CXB), the secondary electrons and positrons rejection efficiency is a bit lower. This is because the behavior of the leptons inside the gas are the same, no matter if they are photoelectrons or cosmic electrons. Therefore the rejection methods are less efficient for low energy electrons and positrons.

\begin{table}[h]
\caption{The residual rates after applying the rejection methods for all the background components.}
\centering
\begin{tabular}{ccc}
\toprule
\textbf{Component}	& \textbf{Residual rate [s$^{-1}$]}	& \textbf{Reject efficiency [\%]}\\
\midrule
Cosmic X-ray		&	5.77E-04	&	66.67\\
Albedo Gamma		&	2.52E-04	&	79.61\\
Albedo Neutron		&	5.24E-05	&	82.35\\
Primary Proton		&	7.86E-03	&	75.12\\
Primary Electron	&	5.52E-05	&	76.89\\
Primary Positron	&	4.39E-06	&	77.00\\
Primary Alpha		&	1.25E-03	&	88.59\\
Secondary Proton	&	2.02E-03	&	85.56\\
Secondary Electron	&	2.84E-03	&	74.35\\
Secondary Positron	&	1.13E-02	&	66.50\\
\midrule
Total				&	2.62E-02	&	75.03\\
\bottomrule
\end{tabular}
\label{table:bkgreject}
\end{table}

\begin{figure}
\centering
\subfloat[]{
\includegraphics[width=0.9\columnwidth]{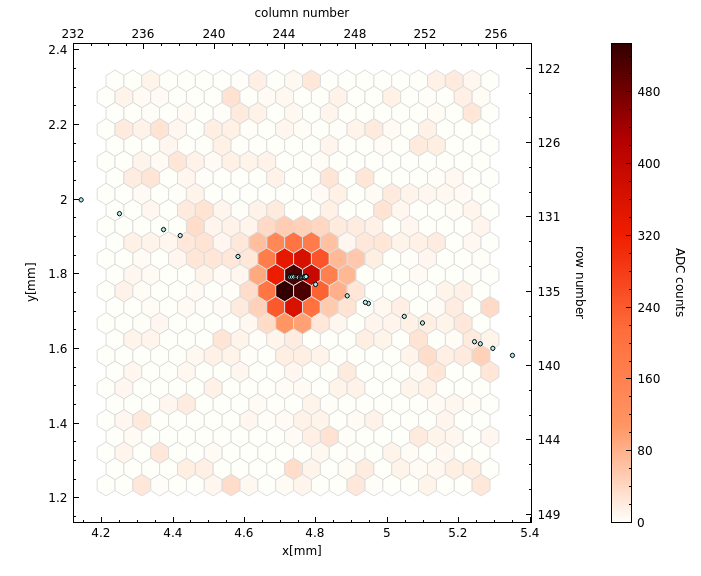}}
\hfill
\subfloat[]{
\includegraphics[width=0.9\columnwidth]{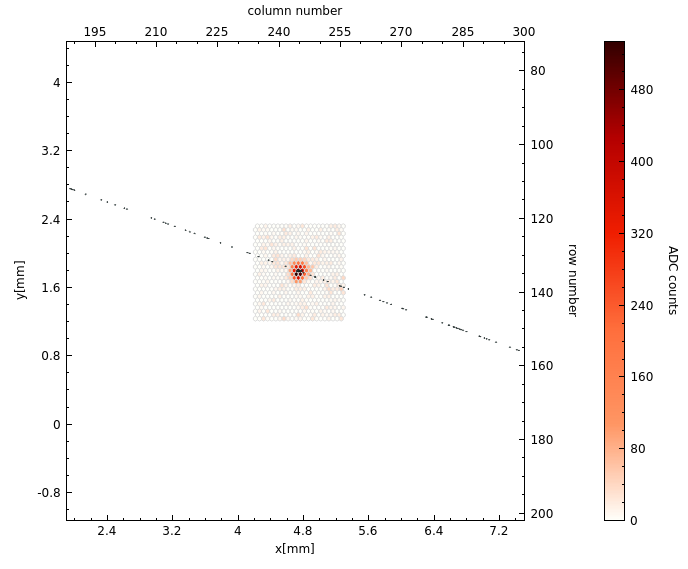}}
\captionsetup{width=1\linewidth}
\caption{A residual background event from an energetic cosmic ray proton with energy deposition of 2.13~keV from the simulation. {\em (a)\/} represents the ROI with window size of 572 pixels,  {\em (b)\/} is the zoomed out view to illustrate the track of the MIP better. Notice that the long and straight track out of the ROI from {\em (b)\/} is still inside the active volume just not triggering. Random electronic noise are added and the green dots present the projected energy deposited positions as Fig.~\ref{fig:trackcmp}.}
\label{fig:deltaray}
\end{figure}

The geometric shape of a typical track from a MIP penetrating the gas cell is easy to be recognized, but in the close encounters, electrons with sufficient kinetic energies may be generated. These electrons, called delta rays, have energy of a few keV and are able to induce further ionization. The tracks from delta rays are similar to those of photoelectrons, become the main contributor of the background events, which is seen in our simulation. 
For example, Fig.~\ref{fig:deltaray} is a residual background event from an energetic cosmic proton, (b) is the zoomed out view of (a). Trigger is generated by a delta ray with high-density energy deposition in the center of the ROI (see Fig.~\ref{fig:deltaray} (a)), while it's mother particle passes through the active area and generates a straight but less dense track nearby (see Fig.~\ref{fig:deltaray} (b)). 
We see from the simulation that primary tracks arriving quasi parallel to the readout plane do not trigger. Indeed if this was the case ROI should be much larger. 
It means that the energy density deposited from the MIP in this case is below the threshold.

\begin{figure}
\centering
\includegraphics[width=\columnwidth]{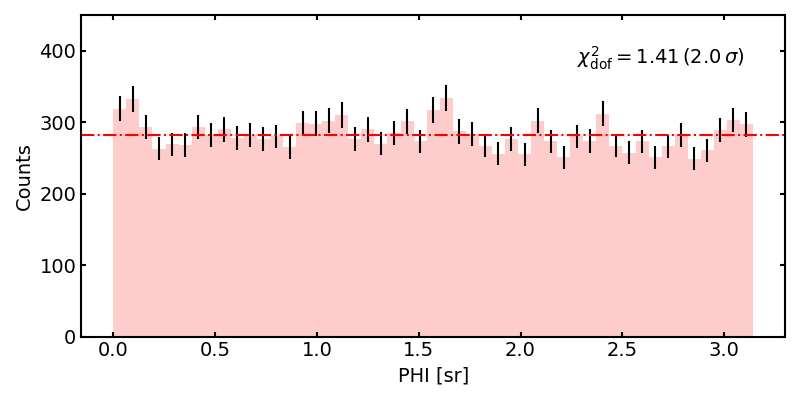}
\caption{The reconstructed angle distribution from the track analysis of the background events. The dashed line is the result from the constant-fit of the histograms.}
\label{fig:bkgphi}
\end{figure}

Fig.~\ref{fig:bkgphi} shows the distribution of the reconstructed angles from the background tracks. No preferential direction is presented in the background simulated data, as we would expect, since the active area of the detector is far away from the side walls of the detector. Background may dilute the measured polarization but according to our simulations, it is not introducing any spurious modulation.

2) Then we apply the cuts to a polarized Crab-like point source simulation data (no background is included).  A model of power-law with the photon index of 2.05 and polarization degree of 100\% is applied. Different from the isotropic background, the source is simulated on-axis along the telescope optical axis.
It is worth mentioning here that, as we will discuss in the next section, background rejection is not needed for the observation of point sources. Nevertheless, here we are investigating how such rejection methods will affect data which are representative of a real astronomical source.

Background rejection methods remove 14.9\% events in 2--8~keV from such a Crab-like point source. Further study shows that these removed events mainly have less deposited energy and less skewness, due to the tracks containing less information about the initial photoelectron directions, such as the events converted in the Al coating of the Be window or in the top Cu layer of the GEM. Depending on the spectrum of the particular celestial source under study, these events may be removed to improve the quality of polarization measurement. The possible selections are consistent with the background rejection methods, and this will reduce the impact of the latter. We will see in the following paragraphs that simulations of the faint extended source suggests that background subtraction is needed to improve the quality of polarization measurement.

The modulation factors before and after applying the rejection methods in three bins are list in Table~\ref{table:mod}.
We find that the modulation factors may change no more than a few percent, for example, there is a relative variation of 2.7\% for Bin 1. 
Nevertheless, the difference are barely consistent from the statistical point of view (at 2.955 $\sigma$), and we then conclude that the impact of the rejection technique on the modulation factor is minor, if any.
This is important because, when a selection of data has been applied, strictly speaking the response function of the instrument has been changed and new one has to be generated, at least when changes are substantial.

\begin{table}
\caption{Modulation factor before and after applying the rejection methods on assumed 100\% polarized source.}
\centering
\begin{tabular}{cccccc}
\toprule
\textbf{Energy bin}	& \textbf{M before rejection}& \textbf{M after rejection}\\
\midrule
Bin 1		& 	25.10 $\pm$ 0.15	&	24.43 $\pm$ 0.17	\\
Bin 2		& 	43.69 $\pm$ 0.29	&	43.43 $\pm$ 0.32	\\
Bin 3		& 	55.44 $\pm$ 0.57	&	56.27 $\pm$ 0.61	\\
\bottomrule
\end{tabular}
\label{table:mod}
\end{table}

%%%%%%%%%%%%%%%%%%%%%%%%%%%%%%%%%%%%%%%%%%
\section{Discussion}  \label{sec:discussion}
In imaging X-ray astronomy, background is expected to influence the observations of very faint sources, depending on various factors, of which the angular resolution is the most important.
Below a certain level of source luminosity, the fluctuations on the background rate prevail on the source rate and the observation becomes background limited. 
The angular resolution of IXPE, in terms of half-energy width (HEW), defined as containing half of the counts from a certain direction, is $\ang{;;27.4}$ (full opening angle). Projected onto the readout plane, the HEW is a round area of $\sim$ 0.22~mm$^{2}$ with the mirror focal length of 4~m.
For point-like sources the source signal to background ratio is relatively high enough, while for extended source this ratio is usually significantly smaller and has to be carefully evaluated.
Here we discuss how these considerations apply to two cases of very high interest for the IXPE program.
We present the results of the simulated observations for a very faint point source, the magnetar AXP 1RXS J1708-4009 (flux $\sim 4.3\times10^{-11}$ erg\,cm$^{-2}$\,s$^{-1}$ in 2--8~keV) , and for an extended source, the Tycho supernova remnant (flux $\sim 1.6\times10^{-10}$ erg\,cm$^{-2}$\,s$^{-1}$ in 2--8~keV), using background levels derived from this paper (both unrejected and residual level in the 2--8~keV energy range).

The observation simulations are performed with IXPEOBSSIM, a simulation framework specifically developed for the IXPE mission (see \citep{2019NIMPA.936..224P} for details). 
IXPEOBSSIM is a Python-based Monte Carlo framework that, takes the source models (including morphological, temporal,
spectral and polarimetric information) as input, convolves them with the analytic instrumental response functions (i.e. the effective area, the energy dispersion, the point-spread function and the modulation factor), produces output files in a format widely used in X-ray community. 
Chandra images are supported as a model for the source morphology. 

As the internal background is already defined in units of counts\,s$^{-1}$\,cm$^{-2}$\,keV$^{-1}$, there is no need to convolve this spectrum with the response function of the optics or the efficiency of the detector.
According to Fig.~\ref{fig:bkgphi}, background is assumed to spread on the readout plane evenly, polarization fraction is zero and polarization angle is uniformly random in [0, 2$\pi$].  

Three separate cases are studied for comparison for both targets (1RXS J1708 for short and Tycho SNR):
\begin{itemize}
\item Case 1: Source-only
\item Case 2: Source plus the residual background (the spectrum as seen in Fig.~\ref{fig:bkgspecreject}, top panel, in red)
\item Case 3: Source plus the unrejected background (the spectrum as seen in Fig.~\ref{fig:bkgspecreject}, top panel, in blue)
\end{itemize}

For each case, 1000 independent runs with exposure of 1 Ms, a typical observation duration planned for IXPE, have been done.
Data analysis processes on point source and extended source are a bit different, which will be discussed more in the following sections.

\subsection{Simulated observation of a point source}
\subsubsection{Model parameters of AXP 1RXS J1708-4009}
Magnetars are isolated, pulsing magnetic-powered neutron stars with extremely strong magnetic fields.
IXPE will enable the direct measurement of magnetic field strength and geometry, as well as investigation of the quantum-electrodynamic (QED) effect of vacuum birefringence that is detectable only in super-strong magnetic fields (\citep{2014MNRAS.438.1686T}).
As one of the brightest and the few potentially interesting magnetars, 1RXS J1708 is taken as an example here.

\begin{figure}
\centering
\includegraphics[width=\columnwidth]{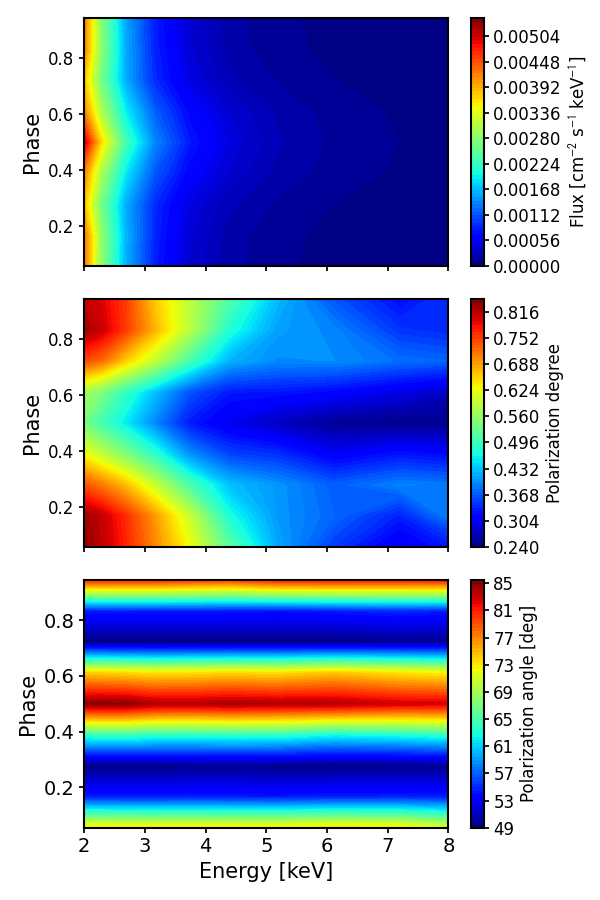}
\caption{1RXS J1708 source model (\citep{2014MNRAS.438.1686T}). From the top to the bottom:  source flux, polarization degree and polarization angle as functions of energy and phase.}
\label{fig:J1708model}
\end{figure}

From \citep{2014MNRAS.438.1686T}, we see that a simultaneous fit of the polarization degree, polarization angle, and phase-dependent flux recovers the input model accurately.
Fitted geometric parameters can also be compared with those inferred from observation of hard X-rays and model fits based on a new
coronal-outflow model. 
IXPE can readily discriminate between models with and without vacuum polarization, thus allowing confirmation of this remarkable QED prediction.
Following the model introduced in \citep{2014MNRAS.438.1686T},  the magnetospheric parameters $\Delta \Phi_{N-S}$=0.5, $\beta$=0.34 and geometrical angles $\chi$=90, $\xi$=60 are set as primaries of the input model. The phase-dependent flux, polarization degree and polarization angle of the input model are shown in Fig.~\ref{fig:J1708model}. Polarization angle is independent of the energy.

\subsubsection{Results}
When dealing with the imaging of a point source, a selection on the region with the encircled energy fraction (EEF) of 0.9 has been done. Here the EEF is the integral profile of the point-spread function (PSF) described as a Gaussian plus a King function, where EEF($\infty$)=1, and EEF(r)=0.9 refers to a round region with radius of r, and contains 90\% of the total events, more details are found in \citep{2014ApJS..212...25F}. In the specific case of IXPE, this is a round area about 4.76~mm$^{2}$ on the readout plane. 
By considering the PSF, the source detection rate is 0.29 counts\,s$^{-1}$ for 1RXS J1708, while the corresponding residual background is $8.51\times10^{-5}$~counts\,s$^{-1}$ and the unrejected background is $3.33\times10^{-4}$~counts\,s$^{-1}$, resulting to the source to background ratio of 3422 and 873 respectively.

\begin{figure}
\centering
\includegraphics[width=\columnwidth]{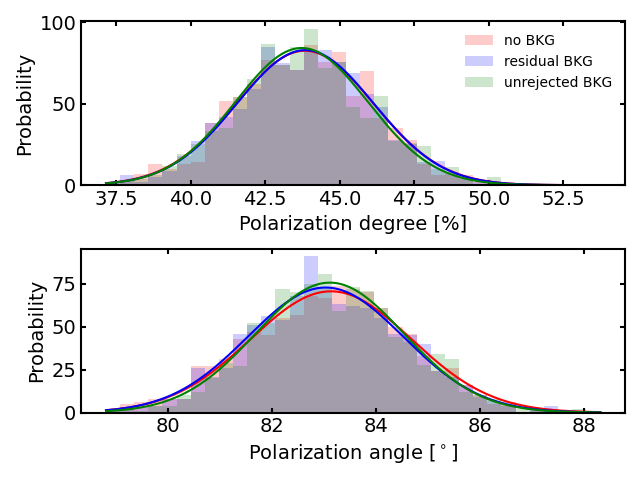}
\caption{The distributions of polarization degrees (top) and polarization angles (bottom) in phase bin [0.44, 0.56] of 1RXS J1708. Red presents the source-only case, blue presents source plus the residual background, and green is for source plus the unrejected background. The histograms are the binned distributions and the solid lines are the best Gaussian fitting. Notice that on the top panel, the red line almost fully overlaps with the blue line.}
\label{fig:J1708result}
\end{figure}

\begin{table}
\caption{Calculated MDPs, and the best fitted values of the polarization degrees and angles in each phase bin of 1RXS J1708. For each phase bin, from the top to bottom, results refer to three cases: source-only, source plus the residual background, source plus the unrejected background. }
\centering
\begin{tabular}{c|ccc}
\toprule
\textbf{Phase bin}	& \textbf{MDP [\%]}	& \textbf{Degree [\%]} & \textbf{Angle [$^{\circ}$]} \\
\hline
\multirow{3}{*}{[0.00, 0.11]	}	&	8.33	& 	73.17 $\pm$ 0.09	&	72.75 $\pm$ 0.04\\	
{}						&	8.33	&	72.85 $\pm$ 0.08	&	72.72 $\pm$ 0.04\\
{}						&	8.34	&	73.11 $\pm$ 0.09	&	72.77 $\pm$ 0.04\\
\hline
\multirow{3}{*}{[0.11, 0.22]}	& 	8.25	& 	72.22 $\pm$ 0.08	&	55.13 $\pm$ 0.03\\	
{}						&	8.25	&	72.30 $\pm$ 0.08	&	55.28 $\pm$ 0.04\\
{}						&	8.25	&	72.23 $\pm$ 0.09	&	55.27 $\pm$ 0.03\\

\hline
\multirow{3}{*}{[0.22, 0.33]	}	& 	8.17	& 	63.58 $\pm$ 0.08	&	50.10 $\pm$ 0.04\\	
{}						&	8.17	&	63.73 $\pm$ 0.08	&	49.97 $\pm$ 0.04\\
{}						&	8.17	&	63.70 $\pm$ 0.08	&	50.05 $\pm$ 0.04\\
\hline
\multirow{3}{*}{[0.33, 0.44]	}	& 	7.40	& 	50.52 $\pm$ 0.08	&	64.63 $\pm$ 0.04\\	
{}						&	7.40	&	50.41 $\pm$ 0.08	&	64.64 $\pm$ 0.04\\
{}						&	7.40	&	50.38 $\pm$ 0.08	&	64.67 $\pm$ 0.04\\
\hline
\multirow{3}{*}{[0.44, 0.56]}	& 	6.78	& 	43.82 $\pm$ 0.07	&	83.14 $\pm$ 0.05\\	
{}						&	6.78	&	43.85 $\pm$ 0.07	&	83.03 $\pm$ 0.05\\
{}						&	6.78	&	43.69 $\pm$ 0.07	&	83.12 $\pm$ 0.05\\
\hline
\multirow{3}{*}{[0.56, 0.67]	}	& 	7.33	& 	46.31 $\pm$ 0.08	&	73.04 $\pm$ 0.05\\	
{}						&	7.33	&	46.13 $\pm$ 0.07	&	73.07 $\pm$ 0.05\\
{}						&	7.33	&	46.01 $\pm$ 0.08	&	73.07 $\pm$ 0.05\\
\hline
\multirow{3}{*}{[0.67, 0.78]	}	& 	8.16	& 	62.62 $\pm$ 0.08	&	50.62 $\pm$ 0.04\\	
{}						&	8.15	&	62.52 $\pm$ 0.08	&	50.62 $\pm$ 0.04\\
{}						&	8.16	&	62.51 $\pm$ 0.08	&	50.53 $\pm$ 0.04\\
\hline
\multirow{3}{*}{[0.78, 0.89]	}	& 	8.31	& 	70.73 $\pm$ 0.08	&	55.62 $\pm$ 0.04\\	
{}						&	8.31	&	70.56 $\pm$ 0.08	&	55.64 $\pm$ 0.04\\
{}						&	8.31	&	70.51 $\pm$ 0.09	&	55.68 $\pm$ 0.03\\
\hline
\multirow{3}{*}{[0.89, 1.00]	}	&	8.41	& 	71.09 $\pm$ 0.09	&	79.19 $\pm$ 0.04\\	
{}						&	8.41	&	71.00 $\pm$ 0.09	&	79.25 $\pm$ 0.04\\
{}						&	8.41	&	70.88 $\pm$ 0.08	&	79.33 $\pm$ 0.04\\
\bottomrule
\end{tabular}
\label{table:J1708result}
\end{table}

In order to perform the phase-resolved observation, data are collected in nine, equally divided phase bins. For each run and each phase bin, a polarization degree and a polarization angle are derived, and a MDP (see Equation~(\ref{equ:mdp})) is calculated. For example, the results from the fifth phase bin [0.44, 0.56], which has the lowest polarization degree from the model (see Fig.~\ref{fig:J1708model} middle panel), are shown in Fig.~\ref{fig:J1708result}. 
In such a worst case, polarization degree has a relative change less than 0.05\% when the residual background is added, and this increases only to 0.25\% if the unrejected background is added. Fitting results from all the phase bins referring to these three cases are listed in Table~\ref{table:J1708result} in sequence. We see that MDPs show no difference with background added, and polarization degrees and angles are consistent in all the phase bins, as expected. The conclusion is that the impact of background on point sources, even the faintest ones, is negligible. 
In the case of point sources, background rejection is not needed, and therefore all the source counts could be kept for further data analysis.

\subsection{Simulated observation of an extended source}
\subsubsection{Model parameters of Tycho SNR}
The Tycho SNR is an extended, relatively faint object that is part of the IXPE observing plan. 
The interest in the source lies in the possibility of observing a high polarization fraction (up to $\sim$50\%, \citep{2011ApJ...735L..40B}) due to synchrotron emission in the shock regions. Such observations would help clarify cosmic rays acceleration processes and test nonlinear diffusive shock acceleration model (see for example \citep{2011ApJ...735L..40B, 2020ApJ...899..142B}).

\begin{figure}[]
\centering
\includegraphics[width=0.9\columnwidth]{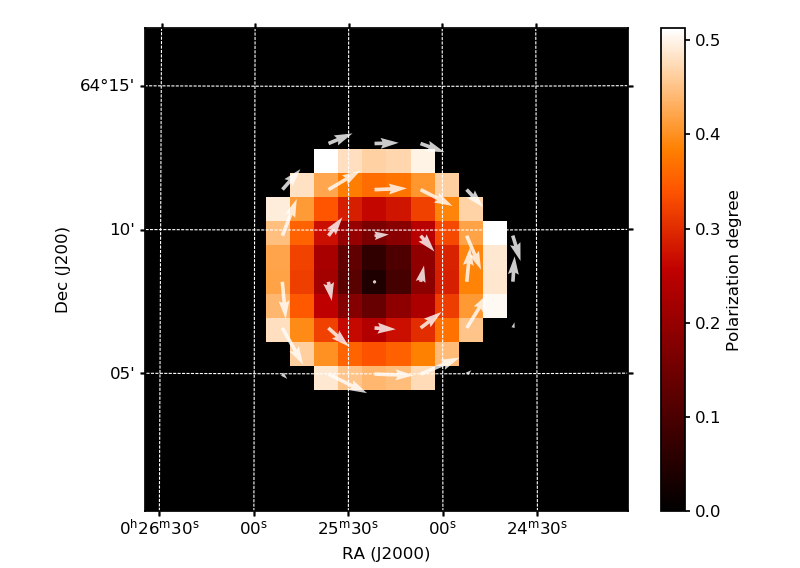}
\caption{Tycho SNR polarization map paired to the source model. The direction and the length of the arrows stand for polarization angle and degree respectively. The maximum polarization degree on the external rim is up to 50\%.}
\label{fig:tycho_model}
\end{figure}

We use a 147~ks Chandra observation (obs ID 15998) as the starting point for the Tycho SNR spectral and morphological model.
Software ixpeobssim takes care of the conversion from the Chandra reprocessed data, an event list includes the energetic, spatial and temporal information, into a corresponding IXPE event list, through effective area correcting, down sampling and events smearing with the instrument response functions. 
A simple polarization model which is a geometrical, radially-symmetric pattern with a maximum polarization degree of 50\% on the SNR external rim, is paired to the source model, as shown in Fig.~\ref{fig:tycho_model}.
The real measured polarization degree is expected to be lower than the model because of the mixing with the unpolarized emission from the thermal component of the supernova remnant. However, this toy model fulfills the purpose of this work, that is to evaluate the relative effect of the instrumental background on the measurement of polarization.

\begin{figure}
\centering
\includegraphics[width=0.8\columnwidth]{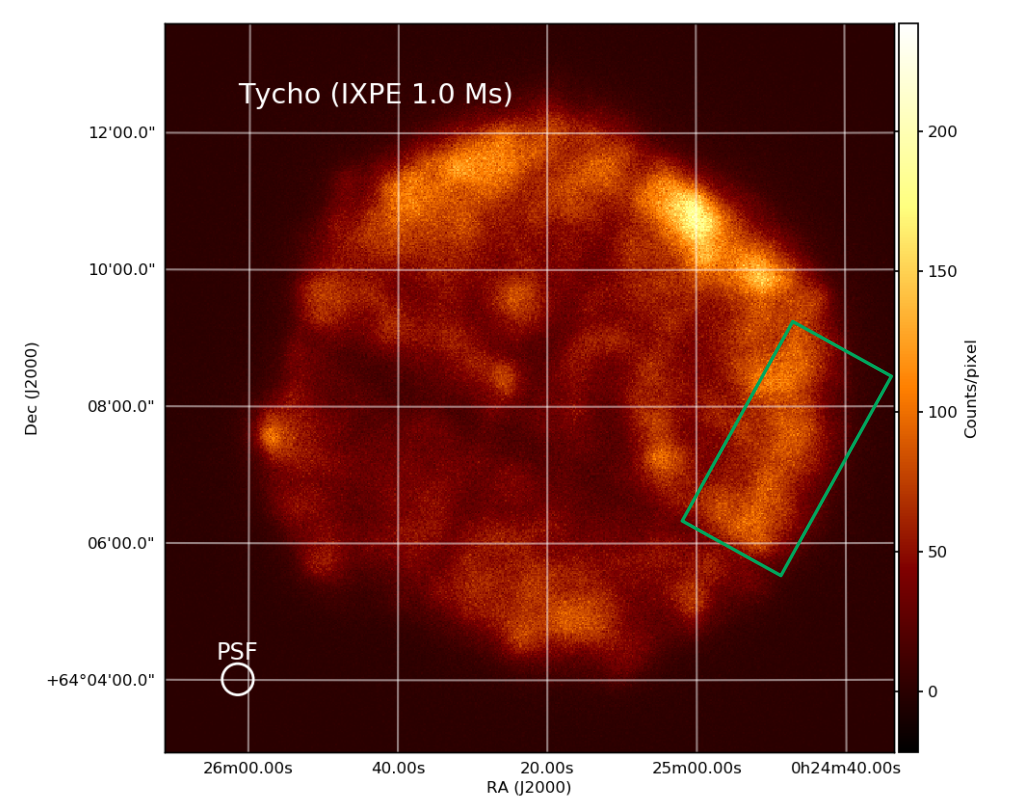}
\caption{The simulated image of the Tycho SNR with 1~Ms exposure derived from ixpeobssim. The rectangle in green is the region selected for background study.}
\label{fig:tychoixpe}
\end{figure}

\subsubsection{Results}
Fig.~\ref{fig:tychoixpe} shows the simulated image of the Tycho SNR with our simulator, the rectangle of $\ang{;;100}\times\ang{;;200}$ is the region we selected for the further data analysis. This region includes the non thermal "stripes" features (\citep{2011ApJ...728L..28E}) that are expected to be highly polarized (\citep{2011ApJ...735L..40B}).

\begin{figure}
\centering
\includegraphics[width=\columnwidth]{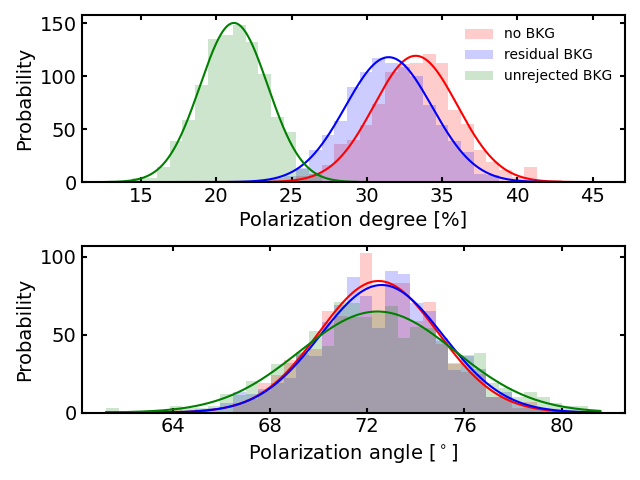}
\caption{The distributions of polarization degrees (top) and polarization angles (bottom) in the Tycho selected region. Red presents the source-only case, blue presents source plus the residual background, and green is for source plus the unrejected background. The histograms are the binned distributions and the solid lines are the best Gaussian fitting.}
\label{fig:tychoresult}
\end{figure}

Fig.~\ref{fig:tychoresult} presents the results of the selected region across the 1000 independent simulations for the three cases under consideration, and the Gaussian-fitted results are listed in Table~\ref{table:Tychoresult}.
For the selected region, the source rate is $1.10\times10^{-2}$~counts\,s$^{-1}$, the residual background is $6.67\times10^{-4}$~counts\,s$^{-1}$,  and the unrejected background is $6.23\times10^{-3}$~counts\,s$^{-1}$. Therefore the corresponding source to background ratio are 16.6 and 1.77, much less than in the case of point sources.
Obviously adding the background have the effect of diluting the measured polarization degree. The uncorrected modulation amplitude for Case 2 in average is smaller by a factor of 5.5\% with respect to the case with no background. This factor dramatically increases to 44.4\% for Case 3.
Though no significant shiftings on the polarization angle, a widen effect is clearly seen for Case 3.
Background subtraction is recommended for extended faint sources. Background data for subtraction can be acquired either as blank-sky or Earth occultation or with the filter and calibration wheel in closed position (\citep{2020JATIS...6d8002F}).

\begin{table}
\caption{Calculated MDPs, and the best fitted values of the polarization degrees and angles for the selected region of Tycho SNR. Results refer to three cases: source-only (Case 1), source plus the residual background (Case 2),  source plus the unrejected background (Case 3). }
\centering
\begin{tabular}{cccc}
\toprule
{}	& \textbf{MDP [\%]}	& \textbf{Degree [\%]} & \textbf{Angle [$^{\circ}$]} \\
\midrule
\textbf{Case 1}	&	8.81	& 	33.25 $\pm$ 0.09	&	72.46 $\pm$ 0.08\\	
\midrule
\textbf{Case 2} 	&	9.05	&	31.45 $\pm$ 0.09	&	72.59 $\pm$ 0.08\\
\midrule
\textbf{Case 3} 	&	10.91&	21.17 $\pm$ 0.07	&	72.42 $\pm$ 0.10\\
\bottomrule
\end{tabular}
\label{table:Tychoresult}
\end{table}

%%%%%%%%%%%%%%%%%%%%%%%%%%%%%%%%%%%%%%%%%%
\section{Conclusion} \label{sec:conclusion}
The background of an instrument in orbit can be evaluated through the comparison with previous space missions, as similar as possible to the new instrument, and scaling the data with reasonable criteria. The GPD has many commonalities with proportional counters that have been used many times in X-ray astronomy, especially in early times. Unfortunately none of them was filled with DME.
The major contributors to the background, to be accounted when scaling from one payload to another one, are the orbit and the mass around the detector. The orbit determines the radiation environment. The mass determines how some components of the radiation are absorbed, so reducing the background, and some others are converted producing the reverse effect. 

Given that the background of gas filled counters is very different for different filling gases, in the past the comparison was always processed with measurements by detectors filled with low atomic number gases  flown on the OSO-8 satellite (\citep{1978ApJ...220..261B}), which is also the initial reference for the estimation of IXPE background. It is not intuitive whether background levels from Neon-filled detector or Methane-filled detector is more representative, Neon has a similar absorption coefficient with the DME, while Methane is more similar in terms of atomic number. 
For the sake of completeness we remind that the detector of ARIEL-6 was filled with Propane (\citep{1983ITNS...30..485M}), likely the most similar molecule to DME, but in literature only data up to 1.5~keV are reported. Anyway at the lower energy range the background seems more similar to that of OSO-8 filled with Neon.
The background suppression methods applied for them were a combination of pulse height discrimination, pulse shape discrimination (PSD) and anti-coincidence veto (\citep{1978ApJ...220..261B}, \citep{1983ITNS...30..485M}), IXPE could, in principle, have better capability to remove background for the following reasons:
(1) PSD is used to identify the charged particles when they produce longer electron tracks than photons depositing the same energy inside the gas. The imaging capability of GPD allows easy discrimination against these events;
(2) Properties derived from the image, related to the shape or the charge density of the track, work as the particle discriminator;
(3) A high ratio of the total gas volume to the active gas volume is beneficial to prevent the background incidence from the side walls, and  the peripheral region of the gas cell can serve as the anti-coincidence veto (\citep{2012SPIE.8443E..1FS}).

\begin{figure}
\centering
\includegraphics[width=\columnwidth]{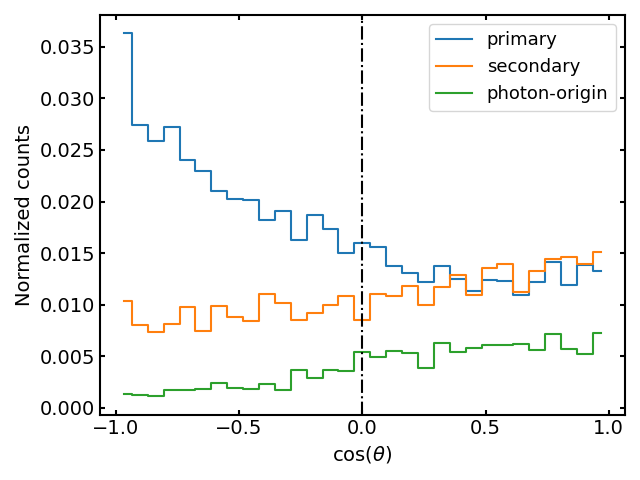}
\caption{The distribution of background incident directions, where $\cos(\theta)$ = 1 presents the direction along the optical axis, and dashed line is $\theta$ = \ang{90}. Blue presents the primary cosmic rays, orange for the secondary components, and green for the photon-origin components.}
\label{fig:parDirection}
\end{figure}

From the Monte Carlo simulation, we understood that the lack of bottom anti-coincidence is responsible for a fair good amount of un-rejected background. Fig.~\ref{fig:parDirection} shows the incident directions of the readout background events, including primary cosmic rays ($\sim$ 5~GeV--100~GeV), secondary cosmic rays ($\sim$ 10~MeV--10~GeV) and photon-origin components (\textless 2~MeV), where $\cos(\theta)$ = 1 presents the direction along the optical axis, and the dashed line ($\theta$ = \ang{90}) illustrates a naive transition from top-coming to bottom-coming direction. For primary cosmic rays, most of the events are from the bottom. This could be explained as the energetic particles interact with the massive materials below the detectors (like the platform), an amount of secondary particles are generated, and they have a certain probability to be detected. If there is bottom anti-coincidence, these events could be removed. Moreover secondary cosmic rays are arriving mostly from the top though, being MIPs (down to $\sim$ 1~MeV for electrons) penetrating the gas cell, they could be discriminated by a back-side veto, too.
We notice that the back side anti-coincidence was usual in wire chambers used in early times. It disappeared for intrinsic reasons in all the subsequent detectors based on a drift region and a sense plane including multi-wire chambers, micro-strip and will not be viable in GPD based mission of IXPE.    

Another significant type of unrejected background events are triggered by delta rays (for example, Fig.~\ref{fig:deltaray}).
The energy of the delta rays is always small compared to the incident energetic particle, so the track of delta rays is usually formed close to the primary track (\citep{2000rdm..book.....K}). However in some cases, primary tracks are not triggering because of the low energy density. 
With GPD what we see is the projection of the tracks.  A particle coming from the direction near the optical axis produces in the image a cluster with a charge density much higher. 
But if an energetic particle coming from the side (\ang{90} off from the optical axis) and crossing the gas, it likely losses energy about 3.2~keV per cm in DME, and in average generates 160 electrons inside the active volume. This means every two holes of GEM along the crossing direction, there is one electron.
Normally such low charge density will not be detected, because of the relatively low gain of the GPD (the level of $\sim$200). 
If there is one electron, with the gain of 200, the amplified charges coming from the same hole spread over 6 to 8 pixels, the average charge collection for each pixel goes below the threshold (noise level of $\sim$50 electrons per pixel). So primary tracks are not detected.
In some sense, that the detector is not triggered by the primary particle is a good thing, but not for removing signals generated by delta rays.
This crucial difference with respect to proportional counters, which usually have a gain of the order of 5000--10000, could be the reason to explain the relatively high background level.

\begin{table}
\caption{Comparison of the residual background levels from detectors aboard OSO-8 and PolarLight with our estimation of IXPE.}
\centering
\begin{tabular}{ccc}
\toprule
\multirow{2}{*}{}		&	\textbf{Energy range}	& 	\textbf{Background}\\
{}				& 	[keV]					&	[s$^{-1}$\,cm$^{-2}$\,keV$^{-1}$]\\
\midrule
\textbf{Methane / OSO-8}		&1.55--3.65	& $6.1\times10^{-4}$\\
\midrule
\textbf{Neon / OSO-8}		&1.6--3.0		& $1.5\times10^{-4}$\\
\midrule
\textbf{Neon / OSO-8}		&3.0--6.0		& $1.0\times10^{-4}$\\
\midrule
\textbf{DME / PolarLight}		&2.0--8.0		& $1.0\times10^{-3}$\\
\midrule
\textbf{DME / IXPE}			&2.0--8.0		& $1.9\times10^{-3}$\\
\bottomrule
\end{tabular}
\label{table:bkgcmp}
\end{table}

Recently a DME-filled GPD was flown aboard the CubeSat PolarLight. The filling gas was the same used for IXPE, however the difference in spacecraft mass distribution and orbit are very significant.
The PolarLight team published the background result in paper \citep{2021arXiv210106606H} in 2021. In general, their results are in good agreement with ours:
(1) The total background rate in the energy range of 2--8~keV measured by PolarLight is about $2\times10^{-2}$~counts\,s$^{-1}$\,cm$^{-2}$ in the central region. This is consistent with our result of $4.66\times10^{-2}$~counts\,s$^{-1}$\,cm$^{-2}$, considering that many impact factors exist through the whole simulation.  PolarLight works in a nearly polar orbit, and it is expected to suffer a higher flux of background, but from their result, the orbital variation of background is small for this type of detector;
(2) The dominant background for PolarLight is induced by the electrons and positrons (they are discussed as a whole in \citep{2021arXiv210106606H}),  and this is the same for IXPE (if you sum the contributions from electrons and positrons together from Table~\ref{table:bkgrate}). We notice that the absolute values of the dominant component (1.52~counts\,s$^{-1}$\,cm$^{-2}$ for PolarLight vs 1.98~counts\,s$^{-1}$\,cm$^{-2}$ for IXPE) are very close, but their fractions with respect to the total (76\% for PolarLight vs. 43\% for IXPE) are different. This is because the second dominant component (background induced from cosmic rays protons) is higher for IXPE. As we discussed before, background induced by the primary cosmic rays protons most likely come from the bottom of the detector (see Fig.~\ref{fig:parDirection}), through the interaction with the massive materials below. While PolarLight doesn't suffer from it thanks to its light mass;
(3) They reported that 72\% background events in the 2--8~keV energy range could be rejected with an effective algorithm, and this efficiency is 75\% from our work (see Table~\ref{table:bkgreject}). The residual events cannot be discriminated as they are generated by the same process as the photoelectron of a few keV.

The comparison of the backgrounds we have discussed are shown in Table~\ref{table:bkgcmp}, including measurements from Methane-filled detector, Neon-filled detector aboard OSO-8, DME-filled GPD aboard PolarLight, and the Monte Carlo simulated result of IXPE from this work. 
With the background rejection methods developed in this work, we removed 92.6\% of background readout events, 
leaving a residual background level of $1.16\times10^{-2}$~counts\,s$^{-1}$\,cm$^{-2}$ in the 2--8~keV energy range. 
This residual background is still 2.9 times higher than the requirement $4\times10^{-3}$~counts\,s$^{-1}$\,cm$^{-2}$, but the requirement is almost one order of magnitude above the value that can be tolerated when observing the most extended and faintest sources IXPE planned (the X-ray reflected from the Sgr~B2 molecular clouds in the vicinity of the Galactic Center), where the surface brightness is 0.04~counts\,s$^{-1}$\,cm$^{-2}$ per DU. 
We proved that this level of background has no influence on point source observations: all the source counts could be kept without applying the background rejection techniques, which come at the cost of a reduction of the genuine counts from the source.
But the dilution of the polarization degree around a few percent may be important for the faint extended source, and in this case background subtraction is needed.

%%%%%%%%%%%%%%%%%%%%%%%%%%%%%%%%%%%%%%%%%%%%%%%%%%%%%%%%%%%%%%%%%%%%%%%%%%%%
%\begin{acknowledgements} 
We acknowledge the 'Accordo attuativo' ASI-INAF n.2017-12-H.0 for funding the Italian contribution to the IXPE project. We thank Riccardo Campana for sharing the background source spectra. 
%\end{acknowledgements}

%%%%%%%%%%%%%%%%%%%%%%%%%%%%%%%%%%%%%%%%%%%%%%%%%%%%%%%%%%%%%%%%%%%%%%%%%%%%
\bibliographystyle{elsarticle-num}
\bibliography{ixpe}      %% example.bib = bibtex entries copied from ADS
\end{document}